**Meteoritical and dynamical constraints on the growth mechanisms and formation times of asteroids and Jupiter**

Edward R. D. Scott

Hawai'i Institute of Geophysics and Planetology, University of Hawai'i at Manoa, Honolulu, HI 96822, USA
escott@hawaii.edu

Abstract

Thermal models and radiometric ages for meteorites show that the peak temperatures inside their parent bodies were closely linked to their accretion times. Most iron meteorites come from bodies that accreted <0.5 Myr after CAIs formed and were melted by $^{26}$Al and $^{60}$Fe, probably inside 2 AU. Rare carbon-rich differentiated meteorites like ureilites probably also come from bodies that formed <1 Myr after CAIs, but in the outer part of the asteroid belt. Chondrite groups accreted intermittently from diverse batches of chondrules and other materials over a 4 Myr period starting 1 Myr after CAI formation when planetary embryos may already have formed at ~1 AU. Meteorite evidence precludes accretion of late-forming chondrites on the surface of early-formed bodies; instead chondritic and non-chondritic meteorites probably formed in separate planetesimals. Maximum metamorphic temperatures in chondrite groups are correlated with mean chondrule age, as expected if $^{26}$Al and $^{60}$Fe were the predominant heat sources. Because late-forming bodies could not accrete close to large, early-formed bodies, planetesimal formation may have spread across the nebula from regions where the differentiated bodies formed. Dynamical models suggest that the asteroids could not have accreted in the main belt if Jupiter formed before the asteroids. Therefore Jupiter probably reached its current mass >3-5 Myr after CAIs formed. This precludes formation of Jupiter via a gravitational instability <1 Myr after the solar nebula formed, and strongly favors core accretion. Jupiter probably formed too late to make chondrules by generating shocks directly, or indirectly by scattering Ceres-sized bodies across the belt. Nevertheless, shocks formed by gravitational instabilities or Ceres-sized bodies scattered by planetary embryos may have produced some chondrules. The minimum lifetime for the solar nebula of 3-5 Myr inferred from the total spread of CAI and chondrule ages may exceed the median lifetime of 3 Myr for protoplanetary disks, but is well within the 1-10 Myr observed range. Shorter formation times for extrasolar planets may help to explain their unusual orbits compared to those of solar giant planets.

Keywords: accretion, asteroids, Jupiter, meteorites, planetary formation

# 1. Introduction

Our understanding of planet formation—the conversion of a disk of dust and gas into a set of major and minor planets—is based largely on information from asteroids, Kuiper Belt Objects, comet samples and meteorites, solar and extrasolar planets, and dynamical models for planet formation. However, there are numerous problems reconciling information from these diverse sources as studies typically focus on a single stage in a very complex process (e.g., the accretion of embryos into terrestrial planets) or on the formation of a single body at a specific location (e.g., Jupiter). These problems have prevented us from understanding the diversity of planets in our solar system and the dramatically different orbits of extrasolar planets.

The major components in chondrites—chondrules, Ca-Al–rich inclusions (CAIs), metallic Fe,Ni grains and matrix—formed in the solar nebula and accreted into asteroids. They therefore provide our best tangible evidence for the nature of the accreting materials in the inner part of the solar nebula, and the timescales involved in forming the chondritic components and accreting them into planetesimals (Kita et al., 2005). The igneous meteorites—achondrites, irons and stony irons—provide complementary information about the timescales for accretion and melting of differentiated asteroids (Wadhwa and Russell, 2000). However, timescales for accretion and differentiation based on radiometric ages of meteorite samples have provided seemingly contradictory constraints: chondrule ages suggest a prolonged nebula lifetime of several million years, whereas some data for achondrites suggest that the time for accretion and melting was much shorter. These problems have raised questions about the validity of the radiometric ages and stimulated interest in models for forming chondrules in planetary rather than nebular environments (e.g., Wadhwa and Russell, 2000; Kleine et al., 2005).

The formation of the terrestrial planets from dust is thought to have been a hierarchical growth process that occurred in discrete stages. Dust grains settled to the nebula midplane, agglomerating into fluffy fractal aggregates and then km-sized planetesimals in $10^{3-4}$ yr. (e.g., Weidenschilling, 2000; Blum, 2004). Over periods of $10^{5-6}$ yr, these planetesimals accreted to form larger bodies, some of which melted, and eventually tens of Moon-to-Mars sized embryos, which collided to form planets in $10^{7-8}$ yr (Chambers, 2003, 2004). To simplify modeling, it is customary to assume that each stage began when the preceding one finished. This model accounts well for the major properties of the terrestrial planets and the Moon, but it fails to explain how km-sized planetesimals were formed (Youdin, 2004) and it has not been successfully reconciled with the meteorite record (Cuzzi et al., 2005). For example, Weidenschilling et al. (1998) concluded that CAI-chondrule age differences of several Myr are "incompatible with dynamical lifetimes of small particles in the nebula and short timescales for the formation of planetesimals."

For the giant planets, there is added uncertainty about their origin. The planetesimal model suggests that timescales for accreting giant planets (Pollack et al., 1996; Hubickj et al., 2005) are comparable to or greater than the lifetimes of gaseous disks around protostars, which are a few million years in regions of low-mass star formation and less for disks >30 AU in radius in regions of high-mass star formation (Haisch et al., 2001; Bally et al., 2005). This has stimulated interest in the possibility that giant planets, both solar and extrasolar, may have formed in $10^{3-4}$ yr from gravitational instabilities in the protostellar disks (Boss, 1997, 2002; Mayer et al., 2002).

Given Jupiter's size and proximity to the asteroid belt, it seems plausible that the asteroids and meteorites might contain clues to help us to understand how and when Jupiter formed. However, there are currently no firm constraints because of conflicting conclusions about Jupiter's role in the formation and accretion of chondrites and asteroids. Weidenschilling et al. (1998), for example, infer that Jupiter was instrumental in forming chondrules and that chondrites accreted at low velocities after Jupiter formed. But Petit et al. (2001) infer that Jupiter's role was destructive—dynamically exciting the orbits of asteroids and almost emptying the asteroid belt of bodies via resonances. Similarly, Boss (2002) argues that early formation of Jupiter was responsible for limiting the mass of the asteroid belt (currently only ~5 × $10^{-4}$ Earth masses), whereas Bottke et al. (2005a,b) argue that Jupiter's late formation helped to empty the asteroid belt.

Here we use recent constraints from radiometric ages of meteorites and their components, the properties of meteorites and asteroids, and dynamical and thermal modeling for the accretion and evolution of asteroids to help understand how and when the asteroids and Jupiter accreted. We focus on four questions. When did the chondritic components and meteorite parent bodies form? How did the chondritic components accrete into parent bodies? What role if any did Jupiter play in the formation of chondritic components? Did Jupiter form before the asteroids accreted? Finally we address the role of Jupiter in the early evolution of the asteroid belt and discuss implications for the formation of Jupiter and other giant planets.

## 2. When did chondritic components and meteorite parent bodies form?

Dating techniques using long-lived and short-lived isotopes now provide a reasonably consistent chronology for the first ~10 Myr of solar system history based on chondritic components and igneous meteorites that were cooled relatively rapidly (in <$10^6$ yr) and not subsequently modified by impact heating or aqueous alteration (Fig.1). These criteria ensure that radiometric dates provide good estimates of the time of formation or peak metamorphism and that the radiometric clocks were not subsequently reset. Only a very few objects in meteorite collections satisfy these criteria: chondrules and Ca-Al-rich inclusions (CAIs) in very rare, unmetamorphosed and unaltered, type 2 or 3 chondrites, a few metamorphosed type 4 ordinary chondrites, and certain igneous or strongly metamorphosed meteorites—viz., some angrites, eucrites, and acapulcoites. The overall consistency of the ages of these objects inferred from short-lived and long-lived isotopes (excluding very rare, so called FUN-type CAIs, which probably formed very early) provides convincing evidence that the short-lived isotopes, $^{26}$Al and $^{53}$Mn, were adequately mixed in the solar nebula (Zinner and Gopel, 2002; Kita et al., 2005; Sanders and Taylor, 2005). In the case of $^{26}$Al, Mg isotopic data for planets, chondrites, and CAIs suggest that heterogeneities were under ~15% (Thrane et al., 2006), which corresponds to an error of <150,000 yr. The generally consistent chronology in Fig. 1 has helped to alleviate concerns that the short-lived isotopes used for dating were heterogeneously distributed in the solar system because they formed locally through irradiation rather than in a supernova (see Kita et al., 2005; Goswami et al., 2005; Tachibana et al., 2006). For a dissenting view, see Gounelle and Russell (2005) who interpret the relatively small inconsistencies as evidence for large initial isotopic heterogeneities and question the value of the short-lived isotopes as chronometers.

The first objects to form in the solar nebula and survive were CAIs. Their $^{26}$Mg/$^{24}$Mg isotopic variations due to the decay of $^{26}$Al, which has a half-life of 0.73 Myr, show that they probably formed in ≤$10^5$ yr, before any other class of planetary materials (Bizzarro et al., 2004; Thrane et al., 2006). High precision ages based on the decay of long-lived $^{235}$U and $^{238}$U to $^{207}$Pb and $^{206}$Pb, respectively, indicate that CAIs formed 4567.2±0.6 Myr ago (2σ limits; Amelin et al., 2002). [These ages are derived from the $^{207}$Pb/$^{206}$Pb ratio and hence are called Pb-Pb ages (Amelin, 2006).] The refractory nature of CAIs and their short formation period suggest that CAIs formed above 1350 K in the inner part of the solar nebula (<1 AU) when the protosun was accreting rapidly as a Class 0 or I protostar, as these objects have lifetimes of ~$10^4$ and ~$10^5$ yr, respectively (Wood, 2004; Scott and Krot, 2005; Thrane et al., 2006). Thus the formation time of CAIs is a good proxy for the the birth of the solar nebula.

Chondrule ages inferred from $^{26}$Al-$^{26}$Mg and $^{207}$Pb/$^{206}$Pb dating indicate that chondrules formed after CAIs over a much longer period of at least 4 Myr (Fig. 2). Chondrules in CV and CR chondrites formed at 4566.7±1.0 and 4564.7±0.6 Ma, respectively, according to $^{207}$Pb/$^{206}$Pb dating (Amelin et al., 2002, 2004), while the inferred initial $^{26}$Al/$^{27}$Al ratios for chondrules in LL and CO chondrites suggest that both formed 1-3 Myr after CAIs (Kita et al., 2005). Precisely how and where the chondrules and CAIs formed is not clear, but their isotopic variations and the presence of presolar grains in chondrites show that chondrules and CAIs accreted from the solar nebula (e.g., Scott and Krot, 2003, 2005). Each chondrite group contains a mineralogically distinct batch of chondrules, which also tends to be isotopically distinct. Although chondrules in each of the three groups of O chondrites (H, L, and LL) are very similar, chondrules in E, C, and O chondrites have oxygen isotopic compositions that are quite distinct from each other showing that they were formed and accreted in distinctly different nebula locations.

The persistence of CAIs in the nebula for several million years has been regarded as an outstanding problem in meteoritics (e.g., Shukolyukov and Lugmair, 2002). However, Cuzzi et al. (2003) show that outward diffusion due to nebula turbulence can counteract inward drift for periods of more than $10^6$ years. Given turbulent mixing of nebula solids and the lack of mixing between chondrules from different groups, it seems likely that the actual spread of chondrule ages within a group is much less than 1 Myr. In addition, accretion of each chondrite group must have started soon after the formation of its youngest chondrules to minimize mixing of chondrules from different groups.

Chondrules with the youngest formation ages are found in CB chondrites and have Pb-Pb ages of 4562.7±0.5 Myr ago, i.e., they formed 4.5±0.8 Myr after CAIs (Krot et al., 2005). This age is supported by $^{182}$W variations due to the decay of $^{182}$Hf (8.9 Myr half life) that imply that CB chondrites formed 4.9±2.5 Myr after CAIs (Kleine et al., 2005). The origin of the chondrules and metal grains in CB chondrites is less certain as they have very distinctive properties and clearly formed under unusual circumstances, possibly in a planetary impact (Krot et al., 2002, 2005; Rubin et al., 2003). Nevertheless, the presence in CB chondrites of CAIs and highly altered matrix-rich inclusions, as in most other chondrites, suggests that the CB chondrules and metal grains accreted from the nebula when it was still turbulent and capable of mixing materials that formed at diverse nebular locations and times (Cuzzi et al., 2003).

Although the basic chronology shown in Figs. 1 and 2 is fairly robust, we can anticipate minor changes as better samples are analyzed with improved techniques. Some changes may even exceed the quoted statistical uncertainties because an assumption made in deriving the formation age was incorrect, e.g., a radiometric clock may have been reset by alteration or reheating. For example, Baker et al. (2005) derived a Pb-Pb age of 4566.2±0.1 Myr for two angrites and claim from Mg isotopic data for several angrites that the angrite parent body melted 3.3-3.8 Myr after CAIs formed. In this case, CAIs would be ~2.5 Myr older than Amelin et al. (2002) concluded, and the data of Amelin et al. (2002, 2004) in Fig. 2 would shift down by 2.5 Myr relative to the ages derived from the Al-Mg chronometer, which would be unaffected. This change would increase the duration of chondrule formation by ~2.5 Myr, but would not otherwise affect the conclusions in this paper.

If chondrites had formed from the first generation of planetesimals, as the standard model for accretion implies, one could infer that there was insufficient $^{26}$Al to melt planetesimals (Kunihiro et al., 2004), and that the onset of planetary formation was delayed until ~1 Myr after CAIs formed (Kita et al., 2005; Cuzzi et al., 2005). However, the evidence for isotopic homogeneity discussed above and other arguments suggest that the heat source for melting was $^{26}$Al and, to a lesser extent, $^{60}$Fe (see McSween et al., 2002). Thermal modeling shows that with these heat sources, the differentiated bodies must have accreted before the parent bodies of the chondrites and <1-1.5 Myr after CAIs formed (Sanders and Taylor, 2005; Haack et al., 2005). Only bodies <20-30 km in diameter, which are smaller than the parent bodies of the meteorites, and those rich in ice (Wilson et al., 1999) could have accreted at this time without subsequently melting their silicates (Hevey and Sanders, 2006).

The conclusion that igneous meteorites probably come from bodies that accreted before the chondrite parent bodies and were melted by $^{26}$Al has been dramatically confirmed by $^{182}$Hf-$^{182}$W dating of CAIs, chondrites, and iron meteorites (Kleine et al., 2005). Data for the major iron meteorite groups, IIIAB, IVA, and IIAB, show that metallic Fe,Ni separated from silicate to form molten metallic cores in their parent bodies ≤1.5 Myr after CAIs formed (Scherstén et al., 2006; Markowski et al., 2006). Thermal modeling using $^{26}$Al and $^{60}$Fe as heat sources shows that the parent bodies of these iron meteorites accreted <0.5 Myr after CAIs formed (see Bizzarro et al., 2005). Some tungsten isotopic data were initially interpreted as evidence for separation of molten metal and silicate before CAIs formed and that irons in a single group separated from silicate over periods as long as 10 Myr. However, these isotopic data are now thought to reflect a reduction in the $^{182}$W/$^{184}$W ratios by cosmic rays over periods of up to $10^9$ yr (Markowski et al., 2006).

Two groups of iron meteorites contain silicate inclusions, the so-called non-magmatic groups IAB and IIE, and their W isotopic data suggest they formed 5-15 Myr after CAIs (Markowski et al., 2006; Scherstén et al., 2006). However, this may not be the time at which metal and silicate were initially segregated. Data for Morasko, Seeläsgen, and Cranbourne, which are actually group IAB irons, not IIICD (Wasson and Kallemeyn, 2002), overlap the IIIAB range, suggesting that for group IAB, the range of W isotopic compositions reflects partial equilibration of W isotopes during late impact-induced mixing of early-formed molten metal with silicate (Benedix et al., 2000; Bogard et al., 2005), rather than late impact melting.

Chondrites also contain evidence suggesting that their parent bodies accreted after bodies that were melted. Some achondritic fragments in chondrites are found in regolith breccias and probably formed by impact melting long after accretion. However, a few achondritic fragments are found in non-regolith breccias, formed within a few Myr of CAI formation, and probably accreted with chondrules (Kennedy et al., 1992; Mittlefehldt et al., 1998, pp.168-171). The maximum metamorphic temperatures of chondrite groups are correlated with mean chondrule age, as expected if $^{26}$Al was the predominant heat source and the parent bodies were >20-30 km in diameter. CR and CB chondrites have younger chondrules and reached lower metamorphic temperatures than the CV, LL, and CO chondrites.

Radiometric ages for primitive chondrites, their CAIs and chondrules, and rapidly cooled, igneous meteorites and thermal modeling now provide a consistent picture of planetesimal accretion in the early solar system. Meteorite parent bodies did not begin to accrete simultaneously, as the standard model implies. Chondritic materials were largely formed and accreted over a period that began <0.5 Myr after CAIs formed and lasted for at least 4 Myr. Planetesimals >20-30 km in diameter that accreted inside the snow line <1 Myr after CAIs formed were melted by radioactive heat forming silicate mantles and metallic cores, from which achondrites and irons were derived. Chondrites come from bodies that accreted from 1 to at least 4.5 Myr after CAIs formed. Early-formed chondritic parent bodies such as the parents of the CV and LL chondrites were strongly metamorphosed whereas the parent bodies of the CR and CB chondrites, which accreted several half-lives of $^{26}$Al later, were scarcely heated.

## 3. How did the meteorite parent bodies accrete?

The accretion timescale for meteorites of several Myr inferred from isotopic data is compatible with disk lifetimes inferred from astronomical data (e.g., Podosek and Cassen, 1994) but inconsistent with much shorter timescales inferred for planetesimal formation by, for example, Weidenschilling (1988) and Weidenschilling et al. (1998). This has led several authors to suggest that chondrites formed not in chondritic planetesimals, but on the surfaces of preexisting asteroids that were chondrule-free and possibly differentiated (e.g., Kleine et al., 2005). Weidenschilling et al. (1998, 2001) suggested that chondrules formed by nebula melting of recycled debris followed by accretion onto the surfaces of large chondrule-free asteroids. Sears (1998, 2004) agreed that chondrules were volumetrically insignificant in the solar system and argued that massive impacts on large, volatile-rich, carbonaceous asteroids produced plumes of melt droplets, gas, dust and fragments that gradually deposited chondrules onto the surfaces of the asteroids. Finally, Lugmair and Shukolyukov (2001) and Sanders and Taylor (2005) have suggested that chondrules formed by collisions between molten bodies and accreted onto existing or newly-formed bodies. The major concern here is not so much how chondrules formed, but how they were aggregated into bodies.

There are several arguments against the concept of deriving chondrites from surficial layers on chondrule-free asteroids as many meteorite types come from bodies that appear to have been well sampled because they were scrambled or largely demolished by impacts (Keil et al., 1994; Scott, 2002). Breccias of materials derived from diverse depths are known but very few breccias are mixtures of chondrule-free and chondrule-bearing material (Bischoff et al., 2006). The parent bodies of the ordinary chondrites are well sampled by breccias but do not contain abundant chondrule-free material, which, if present, should have been baked into rocks strong enough to survive atmospheric entry (Scott, 2002). If chondrules were only formed late in nebula history after many planetesimals formed, we might expect to find metamorphosed chondrule-free material from early-formed bodies in meteorite collections. However, strongly metamorphosed chondrite groups like CK chondrites and well-sintered meteorite types like acapulcoites and winonaites, which presumably accreted before CR and CO chondrites, contain some chondrules (Krot et al., 2003). In addition, iron meteorites have depletions of moderately volatile siderophile elements that are inversely correlated with condensation temperature, as in chondrites (Scott, 1979; Palme et al., 1988), and broadly similar oxygen isotopic compositions (Krot et al., 2003). Thus it seems unlikely that chondrules accreted into surficial layers on early-formed planetesimals, and more likely that the chondrites and differentiated meteorites come from bodies that accreted separately from planetesimals composed of similar kinds of ingredients.

If the chondrites accreted into separate bodies over 5 Myr, some process must have segregated early and late accreting materials. Wetherill and Inaba (2000) tested an accretion model in which 0.5 km planetesimals were continuously added during a $10^5$ yr period to a 0.2 AU wide accretion zone at 1 AU. They found that the late-forming planetesimals tended to be accelerated by the larger bodies that accreted early, and were destroyed by mutual collisions at impact speeds far above their escape velocity. Fragments of these disrupted planetesimals were captured by larger bodies or drifted into the sun. Thus chondrite groups like CR and CB that have relatively young chondrules (Fig. 1) could not have accreted close to early-accreting asteroids like the parent bodies of the CV chondrites and the differentiated meteorites. How were different batches of chondritic ingredients accreted and preserved during such a prolonged accretion period?

Given the problems of accreting nebular materials in the vicinity of early-formed bodies, it seems likely that the location where planetesimals formed varied systematically over time. For example, a planetesimal formation region may have swept across the asteroid belt so that the early-formed bodies that melted were located closer to the Sun than the chondrites (e.g., Grimm and McSween, 1993). Supporting evidence for some kind of progressive accretion comes from the distribution of asteroid spectral types. Weakly heated, carbon-rich C asteroids are concentrated in the outer belt, whereas more strongly heated S asteroids, which probably supply ordinary chondrites (Chapman, 2004) and may have been partly melted (e.g., Sunshine et al., 2004; Harderson et al., 2006), are concentrated in the inner belt. To account for the large number of bodies supplying us with iron meteorites (about 90 comprising ~70% of the sampled asteroids; see Scott, 1979; Keil, 2000), the lack of achondrites from their associated silicate mantles, and the dearth of igneously differentiated asteroids especially among asteroid families created in catastrophic impacts, Bottke et al. (2006) have suggested that most iron meteorites actually formed at the location of the terrestrial planets and were subsequently scattered into the asteroid belt by planetary embryos.

Planetesimal formation may have started close to the protosun where orbital periods are shorter and surface densities higher. If surface density varied as $~1/r$, time scales for accretion by collisional growth would be about ten times longer in the asteroid belt than at 1 AU. Thus planetary embryos that formed from km-sized planetesimals in $10^5$ yr at 1 AU might take $10^6$ yr to form at 2.5 AU (Wetherill and Inaba, 2000). However, as noted above, the chondrules in a chondrite group that formed late such as the CR group are very different from those that formed early, like the CV chondrules. Thus, other factors besides radial variations in accretion rates affected the formation of planetesimals in the inner nebula. The absence of chondrule-free, dust aggregates in meteorite collections, except for the carbon-rich CI chondrites, and the abundance of chondrules in the other chondrites suggests that except in the outer belt, chondrule formation was an essential stage in initiating the formation of km-sized planetesimals. Contrary to the standard model (e.g., Blum, 2004), fluffy dust aggregates do not appear to have accreted together to form chondrites (except possibly CI chondrites). Instead, chondrules and dust appear to have been intermittently concentrated at the midplane so that planetesimals were formed from dust-coated chondrules. Chondrules may have been concentrated between turbulent eddies (Cuzzi et al., 2001; Cuzzi, 2004) or adjacent local pressure enhancements (Haghighipour, 2005) and then assembled by collisional growth at the midplane. Alternatively, concentration may have resulted from inwards spiraling followed by gravitational instabilities in a dense midplane layer (Youdin and Shu, 2002). In any case, accretion probably started close to the protosun perhaps because chondrule-forming shocks were more intense (e.g., Wood, 2005), or the surface density of solids was higher. Because of the disruptive effects of large, early-formed bodies (Wetherill and Inaba, 2000), the zone where planetesimals could form and survive may have moved outwards over time. Note that possible effects due to scattering of planetesimals by planetary embryos at ~1AU on planetesimal accretion in the asteroid belt have not been modeled.

If the chondrite groups formed sequentially at ever increasing distances from the protosun, we should expect them to form a single sequence in which chemical properties such as abundances of refractory or volatile elements varied systematically. However, the chondrite groups do not form such a sequence and a more complex accretion model is needed. Given the stickiness of interstellar organics at 220-290 K (Kudo et al., 2002), the increase in the surface density of solids at the ice line by a factor of ~2.2 (Chambers, 2006), and the chondrule-free, carbon-rich nature of CI chondrites and cometary interplanetary dust particles (see Krot et al., 2003; Bradley, 2003), it is possible that accretion was also initiated in the outer solar system, where sticky tar or ice was most abundant. Irrespective of whether km-sized planetesimals formed in this region by gravitational instabilities or collisional growth, some mechanism allowed planetesimals to form beyond 3 AU where chondrules were less abundant.

There is no spectral evidence that ice-bearing asteroids in the outer part of the asteroid belt were strongly heated or differentiated, as phyllosilicates have not been detected beyond 2.9 AU (Rivkin et al., 2002). However, the phyllosilicate spectral features may be masked by carbon, and some early-formed, volatile-rich bodies may have been disrupted by gas release (Wilson et al., 1999). Among meteorites that may have formed in the outer belt and been strongly heated are the CK chondrites, which are matrix-rich and strongly metamorphosed like ordinary chondrites above 800°C (Keil, 2000; Krot et al., 2003). There are also some igneous meteorites that probably formed from precursors rich in volatiles and carbon, which may have been derived from bodies that accreted early in the outer belt. Ureilites, which are achondrites containing 2-6 wt.% C, have oxygen isotopic compositions and bulk compositions strongly suggesting they formed from some kind of carbonaceous chondrite (Mittlefehldt et al., 1998; Goodrich et al., 2004). The IAB irons are also carbon-rich with up to 2 wt.% carbon plus graphite nodules up to 10 cm in size (Buchwald, 1975), although they are linked with winonaites, which contain smaller amounts of graphite. Thus, it is plausible that some planetesimals did accrete early in the outer asteroid belt, i.e., <1.5 Myr after CAIs, and that two accretion zones may have approached one another in the asteroid belt (Fig. 3).

In summary, it seems likely that the chondrites are derived from bodies that formed entirely by the aggregation of chondritic material and are not derived from surficial layers on chondrule-free bodies. The formation times of chondrules and CAIs and the properties of chondrites show that chondrite parent bodies, and possibly all planetesimals inside the CI chondrite location, were assembled intermittently over several million years from separate batches of recently formed chondrules and dust. Consolidation of nebular solids into planetesimals may have been triggered by localized concentrations of chondrules in the inner regions, and ice particles or sticky organics in the outer regions. Late-forming chondrules were not able to accrete onto the early-formed bodies, which gravitationally perturbed one another, or between them, and instead formed new planetesimals far from the disruptive effects of large, early-formed bodies. Accretion may have started close to the protosun and also near the snowline or tarline so that two accretion fronts gradually spread across the asteroid belt (Fig. 3). Early formed bodies melted: many iron meteorites and igneous meteorites probably come from bodies that originated in the terrestrial planet region and were scattered into the main belt (Wasson and Wetherill, 1979; Bottke et al., 2006); a few rare carbon-rich igneous meteorites probably formed in the outer belt. The ordinary chondrites and most carbonaceous chondrites, which have close links to the S and C type asteroids comprising the great majority of main belt asteroids, probably formed subsequently in the asteroid belt.

In the following sections we examine the diverse and seemingly incompatible roles that have been proposed for Jupiter during chondrule formation, asteroid accretion, and the destabilization of the asteroid belt. What are the strongest constraints on Jupiter's role that are compatible with the chronology of asteroid formation inferred from meteorites?

**4. Was Jupiter involved in chondrule formation?**

The abundance of chondrules in chondrites and the inferred existence of chondritic components in nearly all unmelted parent bodies of the meteorites imply that energetic events created large volumes of silicate melt droplets in the inner solar nebula. The source of these energetic events has not been identified. The two prime candidates are nebular shocks and the X-wind or protostellar jet model, in which chondrules were generated at the inner edge of the disk and transported out of the plane of the disk to the asteroidal region (see Ciesla, 2005). Various sources for nebular shocks have been proposed including close encounters with sibling protostars (Bally et al., 2005). Here we focus on models for chondrule formation that involve Jupiter. Two rely on Jupiter directly or indirectly to generate shocks to make chondrule melts; the third invokes impacts between planetary embryos perturbed by Jupiter.

While the nebula was still present, Jupiter could have increased the eccentricities of planetesimals via orbital resonances causing them to travel through the gas at supersonic speeds (Weidenschilling et al., 1998; Marzari and Weidenschilling, 2002). These planetesimals may have produced bow shocks capable of melting millimeter-sized silicate aggregates (Hood, 1998; Hood et al., 2005). However, to match the thermal histories of chondrules requires planetesimals with radii of ~1000 km moving at speeds of ~8 km/s and very low nebula opacities (Ciesla et al., 2004). Even under these conditions, only a few percent of the mass of the nebula solids would have been converted into chondrules. There are also problems accreting chondrules on the surface of a pre-existing asteroid (as discussed above) or into freshly formed bodies in the presence of Jupiter, as Marzari and Weidenschilling (2002) propose (see below).

The second process involving Jupiter in chondrule formation envisages that Jupiter and precursor transient clumps that formed by gravitational instabilities both drove strong shock fronts in the inner disk generating chondrules (Boss and Durisen, 2005). Although Jupiter and precursor clumps are strong candidates for producing shocks, once Jupiter became large enough to open a gap in the nebula, chondrule formation would end (Boley and Durisen, 2005).

The third mechanism was proposed by Krot et al. (2005) to explain the highly unusual CB chondrites, which have chondrules that formed 4.5±0.8 Myr after CAIs. These authors infer that Jupiter caused Moon-to-Mars sized planetary embryos to collide in the asteroid belt at high speed generating an impact plume of melt droplets and vapor from which CB chondrules formed. However, we cannot conclude that Jupiter was present as planetary embryos can excite one another and collide in Jupiter's absence. In that case, the CB chondrules (and those in the related CH group) may simply have accreted into planetesimals at a location where asteroids had not yet formed, but alternative mechanisms are also possible. For example, impact plume products may have accreted in orbit forming a satellite that was later detached during close passage of another body.

If any chondrules did form under Jupiter's influence, as these authors have suggested, then the age of those chondrules would provide a lower limit on Jupiter's age. However, it is uncertain whether any melted particles produced by shocks from Jupiter or from Jupiter-scattered embryos would be able to accrete into asteroids. The relative timing of the formation of Jupiter and the asteroids is addressed in the next two sections.

## 5. Did Jupiter form before the asteroids accreted?

Kortenkamp and Wetherill (2000), who were the first to investigate whether planetesimals could accrete inside 5 AU if Jupiter had already formed, found that Jupiter greatly hindered accretion. However, further studies by Kortenkamp et al. (2001a) showed that several Ceres-sized objects could form in 3 Myr at 2.3 AU after Jupiter and Saturn formed. This result was interpreted by Boss (2002) as support for rapid formation of Jupiter by gravitational instabilities in the disk prior to asteroid formation. However, several arguments suggest that asteroids could not readily accrete after Jupiter reached its current mass.

First, Kortenkamp et al. (2001a) found that accretion beyond 2.3 AU, where most asteroids and most of the mass of the belt are currently located, was severely limited by resonances with the giant planets. Second, in their simulations they fixed the locations of Jupiter and Saturn at 1 AU outside their present positions (i.e., Jupiter at 6.2 AU and Saturn at 10.5 AU), arguing that this would allow for later migration caused by planetesimal scattering. However, numerical simulations of giant planet migration suggest that Jupiter would have migrated by less than 0.5 AU due to planetesimal scattering. Scattering of icy bodies into the Kuiper belt by the giant planets caused Jupiter to migrate inwards by less than 0.3 AU (Hahn and Malhotra, 1999; Tsiganis et al., 2005), while scattering of bodies in the asteroid belt caused a smaller shift (Petit et al., 2001). Similarly, the inwards migration distances of Jupiter inferred from the lack of asteroids between 3.5 and 3.9 AU and the properties of the Hilda asteroids were ~0.2 and ~0.45 AU, respectively (Liou and Malhotra, 1997; Franklin et al., 2004). Thus if Jupiter was ever located at 6.2 AU, it must have been migrating inwards due to angular momentum exchange with the nebula gas (see e.g., Chambers, 2003). Under these conditions, accretion in the asteroid belt would have been even more constrained as Jupiter's internal mean-motion resonances would have swept across the belt as it migrated. In their simulations, Kortenkamp et al. (2001b) found that large asteroids could not form in the main belt if Jupiter migrated by 1-3 AU while the planetesimals were accreting.

If Jupiter formed quickly via gravitational instabilities before asteroids accreted in the main belt, it must have migrated slowly far beyond 6.2 AU for >3 Myr when the asteroids were accreting, and then more rapidly to ~5.5 AU before the nebula was dispersed. However, this scenario appears unlikely, as the rate of migration should decrease with time as the disk mass decreases. Although more modeling of asteroid accretion is needed to address, for example, the possibility that Jupiter initially had essentially zero eccentricity (see Tsiganis et al., 2005) and the damping effects of nebula gas, existing models suggest that Jupiter is unlikely to have formed before the asteroids.

## 6. Role of Jupiter in the early evolution of the asteroid belt.

According to several models for the evolution of the asteroid belt, Jupiter's formation triggered the destruction of the asteroid belt and provided an intense period of impact bombardment. The preferred mechanism for removing most of the mass in the asteroid belt and increasing the eccentricities and inclinations of asteroid orbits is that Moon-to-Mars sized embryos formed in the asteroid belt as well as the inner solar system and that they dynamically excited each other and the smaller bodies. When Jupiter and Saturn reached their current mass, there was a dramatic change in the evolution of the belt due to the combined effects of the embryos and the giant planets, and all the embryos and ~99% of the asteroidal bodies were removed through secular and mean-motion resonances with the two planets (Chambers and Wetherill, 2001; Petit et al., 2001).

Mass was probably lost from the belt at many stages: during accretion, through sweeping resonances when Jupiter migrated through the nebula, when the nebula was lost (Nagasawa et al., 2000), and when the giant planets were scattering planetesimals in the outer solar system (Tsiganis et al., 2005; Gomes, et al., 2005; Strom et al., 2005). However, most of the mass was probably lost through the gravitational perturbations of Jupiter and Moon-to-Mars sized embryos as this mechanism accounts for other key features such as the mixing of asteroid types as well as their orbital excitation (Petit et al., 2001, 2002).

Further constraints on the time of Jupiter's formation come from Bottke et al. (2005a, b) who investigated the collisional evolution of the asteroid belt by developing models that would generate the current size-frequency distribution of asteroids, and the numbers, ages and spectral types of of asteroid families. Bottke et al. infer that the primordial population of asteroids <1000 km in diameter was ~150 × larger than at present and that Jupiter formed 1-10 Myr after this population was dynamically excited by planetary embryos. This model therefore requires that accretion ended in the asteroid belt at least a million years before Jupiter approached its current size and location.

## 7. Further Discussion

The time when Jupiter reached its current mass is best constrained by the formation ages of CAIs and chondrules and dynamic models, which show that the asteroids probably accreted before Jupiter was fully formed. Even if we exclude the CB chondrites, the lower limit for the CAI-chondrule forming period given by the ages of CV CAIs and CR chondrules (Amelin et al., 2002) is 2.5±0.8 Myr. Assuming that 1 Myr or more elapsed between the end of asteroid accretion and Jupiter's formation (Bottke et al., 2005b), Jupiter formed >3.5 Myr after CAIs. This makes it unlikely that Jupiter formed via a gravitational instability in the disk, as this process requires massive disks that are well under 1 Myr old (Boss, 1998). Another reason for embracing the core-accretion model for Jupiter rather than the gravitational instability model is that the minimum lifetime for the solar nebula inferred from the ages of CB chondrules and CV CAIs is 4.5±0.8 Myr. This is more than the median lifetime for protoplanetary disks of 3 Myr, but well within the observed 1-10 Myr range (Haisch et al., 2001), and probably long enough for Jupiter to have formed by core accretion (Hubickj et al., 2005; Chambers, 2006).

If Jupiter accreted from planetesimals that formed <1.5 Myr after CAIs, as many models imply (e.g., Rice et al., 2004), we should expect to find small bodies that accreted ice and were strongly heated by $^{26}$Al. However, spectral evidence is lacking and accretion may have been delayed beyond the snow line. The formation times of the giant planets may also be constrained by the properties of their satellites. Incomplete differentiation of Callisto appears to be consistent with core accretion of Jupiter in several Myr (Canup and Ward, 2002; McKinnon, 2006), although Castillo et al. (2005) argue that Iapetus and Saturn formed 1-1.5 Myr after CAIs. Conceivably, organics were more important in accreting Jupiter than ice, as Lodders (2004) inferred from the composition of Jupiter's atmosphere.

If Jupiter as well as the terrestrial planets formed from planetary embryos, it seems almost inevitable that some Moon-to-Mars-sized embryos formed in the primordial asteroid belt, but by no means certain whether there was adequate time throughout the belt for such large embryos to form. Mars itself may be an embryo from the belt (Lunine et al., 2003), but meteorite evidence for their presence is sparse, except possibly in the CB chondrites (Krot et al., 2005). This may be consistent with the widely inferred history of the primordial belt as embryos were removed dynamically and collisions between embryos were rare (Petit et al., 2001). Several meteorite groups, e.g., E chondrites, ureilites, mesosiderites, and IAB, IIE, and IVA irons, show evidence that bodies 100 km or more in size experienced catastrophic collisions within the first 100 Myr suggesting that the rate of collision and total mass of asteroids were greatly enhanced in the primordial belt (Keil et al., 1994; Scott, 2004; Bottke et al., 2005b).

The discussion above implies that bodies that accreted to form the terrestrial planets melted <1.5 Myr after the disk formed. Since lunar-sized embryos probably formed at 1 AU in $10^{5-6}$ yr (e.g., Wetherill and Inaba, 2001; Chambers, 2004, 2006), it is not implausible that embryos were dynamically exciting differentiated planetesimals in the terrestrial planet region when chondrules were still accreting in the asteroid belt. Rare achondritic fragments in chondrites (Kennedy et al., 1992) may be fragments of such bodies that were ejected by early formed embryos, rather than by Jupiter as Hutchison et al. (2001) inferred.

Assuming Jupiter took >3.5 Myr to form, as argued above, it could not have been the prime source for making chondrules in chondrites. Nevertheless, gravitationally unstable clumps (Boss and Durisen, 2005) or proto-Jupiter may have formed shocks capable of generating chondrules at specific stages in nebular evolution. In addition, chondrules may have formed in the envelopes of embryonic Jovian planets (Nelson and Ruffert, 2005), and shocks may have been generated throughout the inner solar system by Ceres-sized bodies that were scattered by embryos, rather than by Jupiter, as Hood et al. (2005) suggested. Since accretion rates to protostars decrease by a factor of around ten over 4 Myr (see Hartmann, 2005), nebular conditions would have changed dramatically over the chondrule formation period. Thus early and late-formed chondrules probably formed and assembled into planetesimals by quite different processes. The youngest chondrules, which are in CB chondrites, are the most likely candidates for an origin involving large planetesimals, embryos, or even giant protoplanets.

The nearly circular orbits of the giant planets in our solar system are unusual when compared with known extrasolar planets, which have circular orbits inside 0.1 AU and very eccentric orbits beyond 0.1 AU (Marcy et al., 2005). Given the protracted lifetime of the solar nebula (>3-5 Myr), it is possible that these orbital differences reflect relatively rapid formation of extrasolar planets in more massive disks. Under these conditions, accelerated growth due to gravitational instabilities and gravitational interactions between planets and disks causing planetary migration and orbital perturbations are more likely. In addition, close approaches of sibling protostars capable of modifying planetary orbits would be more frequent during the earlier stages of protostellar formation (Reipurth, 2005; Bally et al., 2005). Thus late formation of the gas giants in our solar system may be an important clue to understanding orbital differences between solar and extrasolar systems.

Acknowledgements: This work was partly supported by NASA grant NNG05GF68G to K. Keil. I thank W.F. Bottke and H. Palme for helpful reviews and valuable discussions, and A.P. Boss, F.J. Ciesla, N. Haghighipour, S.J. Kortenkamp, A.N. Krot, I.S. Sanders, and many other colleagues for sharing their insights. This is Hawai'i Institute of Geophysics and Planetology Publication No. XXXX and School of Ocean and Earth Science and Technology Publication No. YYYY.

Figures

Fig, 1. Chronology of early solar system based on ages of CAIs, chondrules in ordinary and carbonaceous chondrites (OC, CC), metamorphosed chondrites (H4), primitive achondrites (acapulcoite), and differentiated meteorites (eucrites, ureilites, and angrites) using Mn-Cr, Al-Mg, and Pb-Pb ages; after Kita et al. (2005) who list data sources. Dashed lines connect data obtained from the same meteorite. Vertical arrows connect the samples used to anchor the Mn-Cr and Al-Mg relative ages to the absolute ages inferred from Pb-Pb dating. The 2σ error bars for the Mn-Cr and Al-Mg ages do not reflect the additional errors in the Pb-Pb ages of the anchors. Other chronological constraints can be derived from the Hf-W and I-Xe isotopic systems. Key: Bish, Bishunpur; Sem, Semarkona.

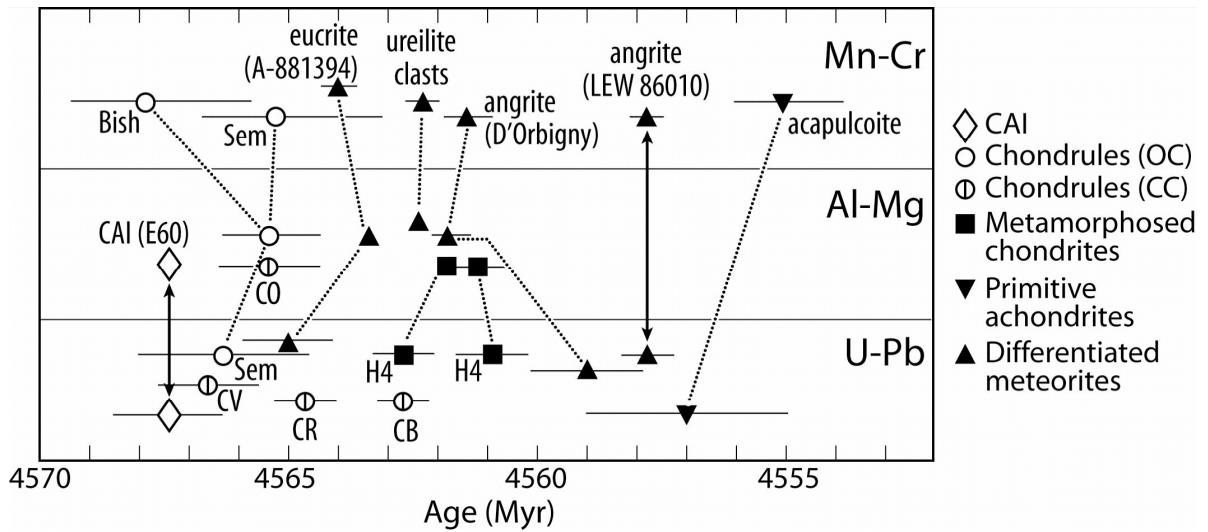

Fig. 2. Ages of chondrules in ordinary and carbonaceous chondrites relative to CAIs inferred from the Al-Mg and U-Pb isotopic systems using an initial $^{26}Al/^{27}Al$ ratio of 5 x $10^{-5}$ and the Pb-Pb age for CAIs of Amelin et al. (2002); diagram after Wood (2005). Both chronometers suggest that most chondrules formed 1-3 Myr after CAIs, although CB chondrules formed 4.5 Myr after CAIs. Sources of data: (1) McKeegan et al. (2000); (2) Kita et al. (2000); (3) Mostefaoui et al. (2002); (4) Amelin et al. (2004); (5) Bizzarro et al. (2004); (6) Kurahashi et al. (2004); (7) Amelin et al. (2002).

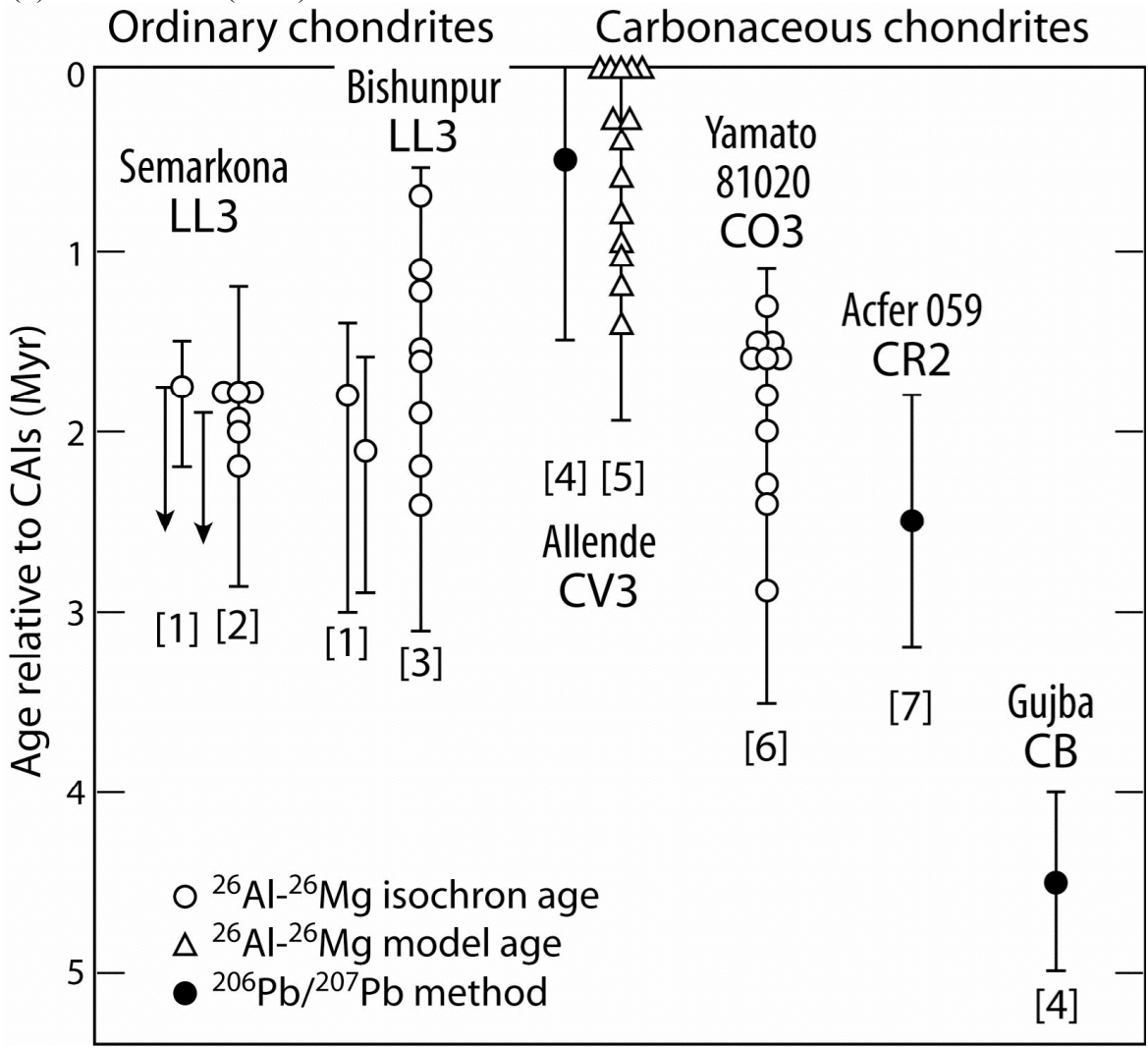

Fig. 3. Schematic diagram showing how the standard model for the accretion of planetesimals and terrestrial planets might be modified to account for the chondrite evidence that planetesimals accreted over a 5 Myr period from various materials formed under diverse conditions. Planetesimals probably began to form near the protosun where dust densities and were highest and chondrules may have been most abundant, and possibly also in cooler regions where sticky organics and ice may have triggered the assembly of carbon-rich planetesimals. Because late-forming materials did not accrete in the vicinity of early-formed planetesimals or onto their surfaces, two accretion fronts (dashed lines) may have gradually approached one another as the nebula cooled. Planetary embryos (larger solid circles) may already have formed and dynamically excited nearby planetesimals on both sides of the asteroid belt when chondrites were accreting within the belt.

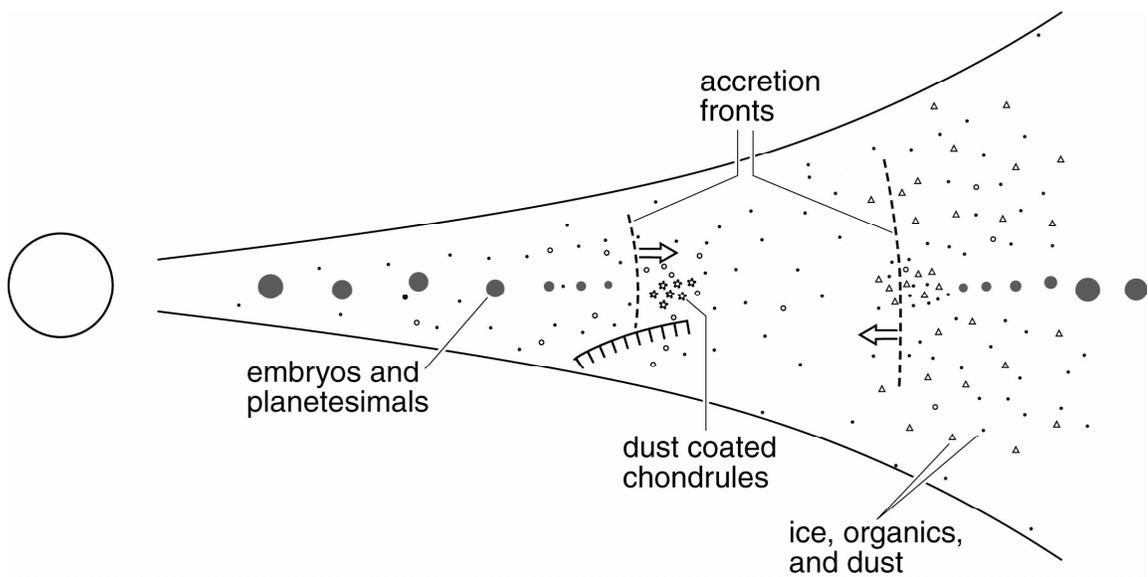